\documentclass[english,twocolumn]{revtex4}

\usepackage[T1]{fontenc}
\usepackage[utf8]{inputenc}
\usepackage{graphicx}
\usepackage{amssymb}
\usepackage{amsmath}
\usepackage{babel}

\newcommand{\bea}{\begin{eqnarray}}
\newcommand{\eea}{\end{eqnarray}}
\newcommand{\beq}{\begin{equation}}
\newcommand{\eeq}{\end{equation}}

\def \ua{{\uparrow}}
\def \da{{\downarrow}}
\def \be{\begin{equation}}
\def \ee{\end{equation}}
\def \ba{\begin{array}}
\def \ea{\end{array}}
\def \bea{\begin{eqnarray}}
\def \eea{\end{eqnarray}}

\def \etal{{\it {et al.}}}

\def \br{{\bf r}}

\def \br{{\bf r}}

\providecommand{\tabularnewline}{\\}

\def \Jw{{J_{w}}}
\def \Js{{J_{s}}}
\def \Jab{{J_{ab}}}

\def \Jws{{J_{w,s}}}

\def \wm{{\omega_{m}}}
\def \ww{{\omega_{w}}}
\def \ws{{\omega_{s}}}

\def \Vs{{V_{s}}}
\def \Vw{{V_{w}}}


\begin{document}


\title{Redistribution of phase fluctuations in a periodically driven cuprate superconductor}
\author{R. H\"oppner$^{1}$, B. Zhu$^{1}$, T. Rexin$^{1}$,  A. Cavalleri$^{2,3}$, L. Mathey$^{1,4}$}

\affiliation{$^{1}$Zentrum f\"ur Optische Quantentechnologien and Institut f\"ur Laserphysik, Universit\"at Hamburg, 22761 Hamburg, Germany\\
$^{2}$Max Planck Institute for the Structure and Dynamics of Matter, Hamburg, Germany\\
$^{3}$Department of Physics, Oxford University, Clarendon Laboratory, Parks Road, Oxford, UK\\
$^{4}$The Hamburg Centre for Ultrafast Imaging, Luruper Chaussee 149, Hamburg 22761, Germany}

\begin{abstract}
We study the thermally fluctuating state of a bi-layer cuprate superconductor under the periodic action of a staggered field oscillating at optical frequencies. This analysis distills essential elements of the recently discovered phenomenon of light enhanced coherence in YBa$_2$Cu$_3$O$_{6+x}$, which was achieved by periodically driving infrared active apical oxygen distortions. The effect of a staggered periodic perturbation is studied using a Langevin and Fokker-Planck description of  driven, coupled Josephson junctions, which represent two neighboring pairs of layers and their two plasmons. In a toy model including only two junctions, we demonstrate that the external driving leads to a suppression of phase fluctuations of the low-energy plasmon, an effect which is amplified via the resonance of the high energy plasmon. When extending the modeling to the full layers, we find that this reduction becomes far more pronounced, with a striking suppression of the low-energy fluctuations, as visible in the power spectrum. We also find that this effect acts onto the in-plane fluctuations, which are reduced on long length scales. All these findings provide a physical framework to describe light control in cuprates. 
\end{abstract}
\maketitle

\section{Introduction}\label{SectInt}
The understanding of high-T$_{c}$ superconductivity in cuprates is one of the central themes in condensed matter physics. While numerous questions about its mechanism and the phase diagram of high T$_{c}$ materials remain, a partial consensus about some of the equilibrium properties of high T$_{c}$ superconductors has emerged, see e.g. Ref. \cite{leggett2006quantum}. The copper oxide planes of these materials are the primary location of the superconducting phenomenon. These planes are weakly coupled in the third direction by tunneling through an insulating layer. A phenomenological description of coupled Josephson junctions is often employed to describe the low frequency electrodynamics for fields perpendicular to the planes. This and similar effective models, such as the XY model, have been discussed in Refs. \cite{bulaevskii1991vortex, bulaevskii1994time, croitoru2012extended, emery1995importance, feinberg1994model, koshelev2013josephson, PhysRev.164.538, van1996transverse, lawrence1971theory, benfatto2008}.

 \begin{figure}
    \includegraphics[scale=0.38]{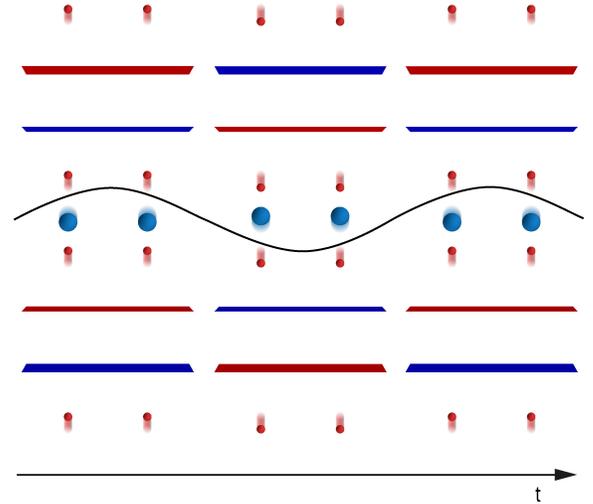}
    \caption{Simplified representation of the crystal structure of YBCO. The copper oxide layers are shown as red and blue, some of the atoms in the insulating layers are shown.  The distortion of this structure, due to the motion of the apical oxygen atoms of the infrared-active B$_{1u}$ mode, discussed  in Ref. \cite{hu2013enhancement},
     is driven periodically in time. This results in an external potential in the CuO layers, that is periodic in time and staggered in the $c$-direction, represented by the red and blue coloring, changing periodically in time.}
    \label{fig:layerstructure}
\end{figure}
 
 A number of recent experiments have explored the dynamical properties of superconducting cuprates, either by analyzing the excitation and relaxation of quasi-particles out and back into the condensate, Refs. \cite{brorson1990, averitt2001, segre2002, giannetti2013}, or by seeking to control the collective properties of the condensate itself with light. This second class of experiments, which involves nonlinear driving  of low energy excitations such as Josephson plasmons and phonons, Refs. \cite{dienst2011bi, fausti2011light, hu2013enhancement, kaiser2012light}, elements of competing order melting and  non-equilibrium phenomena, is  what we study here. 
 
In Refs. \cite{kaiser2012light, hu2013enhancement} an optical phonon mode of YBCO was driven resonantly, enhancing inter-plane coherence and leading to the emergence of  a plasmon edge at temperatures exceeding 300 K, where no signature of superconducting coherence on any time or length scale is observed in equilibrium.
 
  In this paper, we propose a mechanism to reduce phase fluctuations in a layered superconductor, such as YBCO, by driving. We work in an extended, anisotropic XY model, which we drive out of equilibrium. We find that a substantial reduction of the inter-layer phase fluctuations can be achieved under similar conditions as those explored experimentally. This does not only constitute an intriguing scenario of dynamical control in the solid state, it also provides a test for effective theories, such as the XY model, far out of equilibrium.

This paper is organized as follows: In Sect. \ref{Sectmodel} we describe how we represent an optically driven, layered superconductor as an XY model with a driving term. In Sect. \ref{Secttoy} we reduce this model to just two neighboring Josephson junctions, which provides a toy model that displays qualitatively the desired effect of modified phase fluctuations. In Sect. \ref{Sectbulk}, we consider the full three-dimensional system. In Sect. \ref{SectInPlane} we explore the in-plane dynamics of this model, and in Sect. \ref{SectCon} we conclude.

\begin{table}
\begin{centering}
\begin{tabular}{ccccccc}
\hline 
                   & $J_{s}$ & $J_{w}$ & $J_{ab}$ & $E_{c}$ & $A_{0}$ & $\gamma$\tabularnewline
\hline
\hline
 E/$k_{B}$ in K &    $20$       &  $0.2$     &   $100$       &   $6250$ & $20$--$450$       &    $10$\tabularnewline   
\hline
E/$h$ in THz   &    $0.42$    &   $0.0042$  &   $2.1$       &   $130.7$  &  $0.42$--$9.4$    &    $0.2$\tabularnewline
\hline 
E[meV]   &    $1.7$    &   $0.017$  &   $8.6$       &   $539.1$  &  $1.7$--$38.8$    &    $0.9$\tabularnewline
\hline 

\end{tabular}
\par\end{centering}

\caption{\label{tab:parameters} Model parameters in $k_{B} \times$Kelvin, $h \times$THz and meV. All parameters are in energy units, represented by the symbol E in the left column.}
\end{table}

\section{Driven XY model}\label{Sectmodel}
 In this section we develop our model of a driven superconductor. In the experiments reported in  \cite{kaiser2012light, hu2013enhancement}, 
  the optically driven phonon can be seen as a means to periodically modulate the pairing field $\psi_{i}$,  which is the order parameter of the superconducting system, see e.g. Refs. \cite{devereaux}. 
 This order parameter can be written as $\psi_{i} = \sum_{j,k} w(j-i, k-i)\langle c_{j,\ua} c_{k,\da}\rangle$, where $w(j, k)$ is the real space representation of the pairing wave function, located at site $i$. $c_{j, \ua/\da}$ is the fermion operator at site $j$. For a d-wave superconductor, $w(j, k)$ has the corresponding d-wave symmetry.
 
 We consider a situation near the critical temperature $T_{c}$. We assume that at this temperature the bosonic nature of the condensate is not, or only partially, perturbed, that is, we assume that order parameter fluctuations are dominant in destroying superconducting coherence, rather than pair breaking. We also assume that the fluctuations of the field are dominated by thermal fluctuations, which leads us to consider a classical field model, such as the XY model.  
 We approximate the field $\psi_{i}$ in a phase-density representation $\psi_{i} = \sqrt{n_{0} + \delta n_{i}} \exp(i \theta_{i})$, and keep terms up to second in $\delta n_{i}$ in the Hamiltonian. The equilibrium Hamiltonian is  
\bea
H_{0}&=&-\sum_{\langle ij\rangle }J_{ij}\cos(\theta_{i}-\theta_{j})+\frac{E_{c}}{2}\sum_{i}\delta n_{i}^{2}.\label{eq:LD-hamiltonian-eq}
\eea
$\theta_{i}$ and $\delta n_{i}$ are the phase and density fluctuations at site $i$, respectively. $E_{c}$ is the charging energy at each site, i.e. an inverse capacitance. $J_{ij}$ are the tunneling constants between nearest neighbors. There are three tunneling energies: Along the $c$-axis, the values are staggered. $\Js$ represents the strong junctions and $\Jw$ the weak junctions. Within the $ab$-planes the tunneling energy is $\Jab$. 
  We note that the 
 Hamiltonian in Eq. \ref{eq:LD-hamiltonian-eq} extends the standard XY model in two ways. Firstly, since we  investigate dynamics, we added the term $\frac{E_{c}}{2}\sum_{i}\delta n_{i}^{2}$, containing an additional energy scale $E_{c}$. Secondly, the planes connected via $J_{s}$ are  often treated as a single layer. As we describe below, however, introducing the degrees of freedom of these planes is crucial for the mechanism we describe in this paper.
  We also note that the Lawrence-Doniach model introduced in Ref. \cite{lawrence1971theory}, gives rise to inductive coupling between the layers. This type of coupling has been found to be of particular importance to highly anisotropic cuprates, such as BSSCO, while for YBCO the Josephson couplings are the dominant interaction, see Ref. \cite{inductive}. The effect of inductive coupling will be discussed elsewhere.

\begin{figure}
    \includegraphics[scale=0.85]{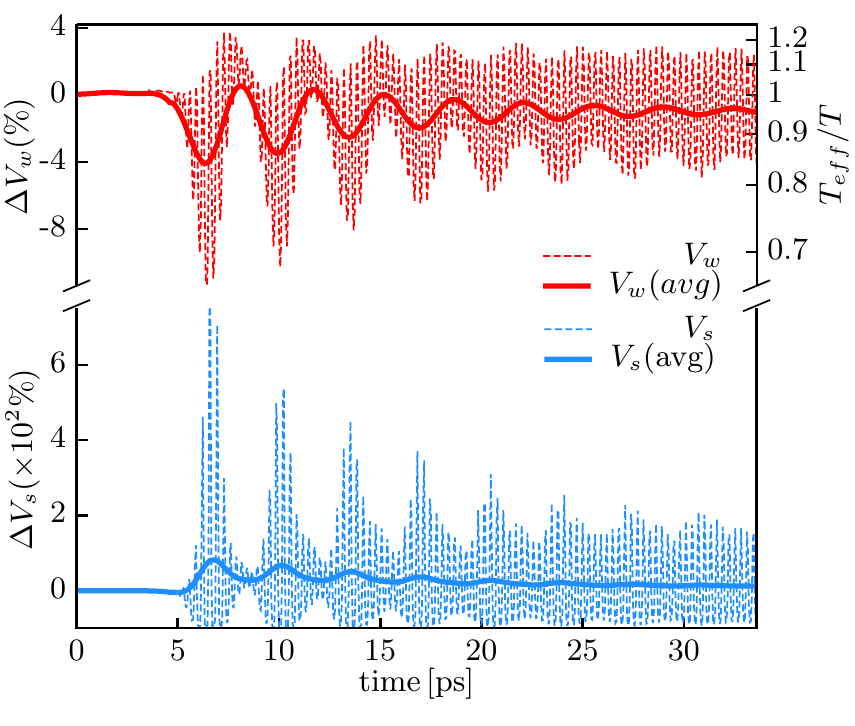}
    \caption{Time evolution of the variances of the weak (red dashed line) and the strong (blue dashed line) junction relative to their equilibrium value, $\Delta V_{w} \equiv V_{w}  - V_{w, th} $ and $\Delta V_{s} \equiv V_{s} -V_{s, th}$, respectively, as percent of their equilibrium values $V_{w, th}$ and $V_{s, th}$, respectively. $V_{w}$ is also represented on the scale of an effective temperature $T_{eff}$ on the right hand side, based on Eq. \ref{Vthermal}.
     The solid lines are  the time evolution  smoothed  via Gaussian averaging with a time scale of $1.8$ ps.}
    \label{fig:timeevolution}
\end{figure}
We model the external driving  with the following term
\bea
H_{dr}&=&\sum_{i} A_{i}(t) \delta n_{i}.\label{eq:LD-hamiltonian-dr}
\eea 
The driving potential  is $A_{i}(t) = (-1)^{z(i)} A_{0} \cos(\wm t)$, where $z(i)$ is the plane index the site $i$ belongs to.
 This describes the effective staggered potential that the electron pairs experience due to the optical phonon distorting the crystal, see Fig.~\ref{fig:layerstructure}. The equations of motion   are
\begin{eqnarray}
\dot{\theta}_{i} & = & E_{c}\,\delta n_{i}+ A_{i}(t)\label{eq:classical-EoM} \\
\delta \dot{n}_{i} & = & -\sum_{j(i)}J_{ij}\sin(\theta_{i}-\theta_{j}).\label{deltan_eq}
\end{eqnarray}
The values of the parameters $\Js$, $\Jw$ and $E_{c}$ are constrained by the two plasmon frequencies of this system. To estimate them, we consider two $ab$-layers coupled either by $\Js$ or $\Jw$. We linearize $\sin\left(\theta_{i}-\theta_{j}\right) \rightarrow \theta_{i}-\theta_{j}$, and diagonalize the system, which gives a gapped and an ungapped dispersion. The gapped dispersion is $\omega_{{\bf k}}^{2} = 2 \Jws E_{c} + \Jab E_{c} (4 - 2 \cos k_{x} - 2 \cos k_{y})$, where ${\bf k} = (k_{x}, k_{y})$ is the lattice momentum, with the lattice constant set to unity.
 We therefore identify $\ww \equiv \sqrt{2 \Jw E_{c}}$ and $\ws \equiv \sqrt{2 \Js E_{c}}$ with the low and high energy plasmon frequency, in the absence of damping. We choose them to be 
 $\sim h \times 1$ THz and $\sim h \times 10$ Thz, respectively, see Refs. \cite{kaiser2012light, hu2013enhancement}. The Kosterlitz-Thouless energy scale of the system is given by $\Jab$, which we choose to be $J_{ab} = k_{B} \times 100$ K. This gives a critical temperature near $100$ K, as for YBCO. 
 The ratios $\Jab : \Js : \Jw$ are approximately of the order of $10^{3} : 10^{2} : 1$.
  This leads to the choice $\Jab = k_{B} \times 100$ K, $\Js = k_{B} \times 20$ K, $\Jw = k_{B} \times 0.2$ K and $E_{c} = k_{B} \times 6250$ K. For the magnitude of the driving potential, we choose  a range of values $A_{0} \approx 2$ -- $40$~meV.  These magnitudes of $A_{0}$ are realistic values, as we discuss elsewhere  \cite{thankSasha}. All the parameters  of our effective model are summarized in Table \ref{tab:parameters}. 

In addition to the Hamiltonian dynamics, we take the coupling to other degrees of freedom into account, such as phonons. We model this by coupling the pairing field to a thermal bath in a Langevin formalism. We extend Eq. \ref{deltan_eq} to
\begin{eqnarray}
\delta \dot{n}_{i} & = & -\sum_{j(i)}J_{ij}\sin(\theta_{i}-\theta_{j}) - \gamma \delta n_{i} + \xi_{i}(t)\label{deltanLE}
\end{eqnarray}
where we added a damping constant $\gamma$ and a classical noise term with $\langle \xi_{i}(t_{1}) \xi_{j}(t_{2})\rangle = (2\gamma T/E_{c}) \delta_{ij} \delta(t_{1} - t_{2})$, where $T$ is the temperature. 
 In this paper we primarily discuss the regime, in which the temperature is below $T_{c}$. Therefore, both plasmon modes are underdamped, and we choose $\gamma = 0.2$ THz. With this choice, the weak plasmon mode is visibly broadened, while still being underdamped, as it should be as the temperature approaches $T_{c}$, while the line width of the strong mode is still fairly narrow.

\begin{figure}
    \includegraphics[scale=0.85]{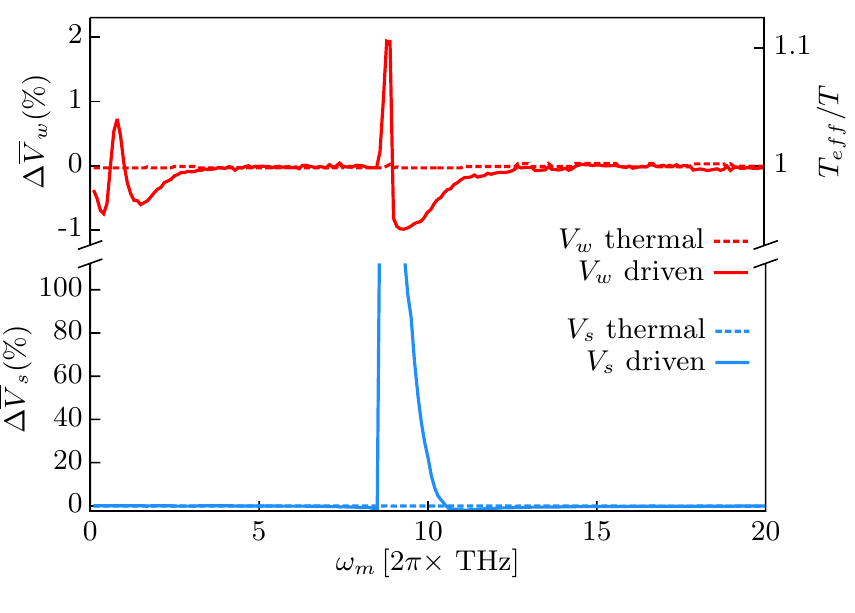}
    \caption{Time averaged variances of the weak junction (upper part, red lines), $\bar{V}_{w}$, and the strong junction (lower part, blue lines), $\bar{V}_{s}$, plotted against the driving frequency. Equilibrium states are shown as dashed lines, and driven steady states as continuous lines.  We find a reduction of $\bar{V}_{w}$ near the resonance of the strong junction, where the strong junction acts as an amplifier of the external driving, and a similar feature near $\omega_{w}$, due to direct driving of the weak junction.}
    \label{fig:V_PS}
\end{figure}

\section{Two oscillator toy model}\label{Secttoy}
To evidence the physical substance of our analysis, it is helpful to first study the problem with a toy model that exhibits only elementary features, nevertheless showing how the driving term in Eq. \ref{eq:LD-hamiltonian-dr} can lead to a reduction of the phase fluctuations of the weak junction.
 First, we ignore the spatial extent of the system in the $ab$-planes. This reduces the system to a 1D system with staggered values of tunneling, $\Js$ and $\Jw$.
 
 We then consider only two neighboring planes, which means we consider three degrees of freedom, $\theta_{i-1}$, $\theta_{1}$ and $\theta_{i+1}$. 
   We assume that $J_{i,i+1} = \Js$ and $J_{i-1,i} = \Jw$, and define the phase differences $\theta_{s} \equiv \theta_{i+1} - \theta_{i}$ and $\theta_{w} \equiv \theta_{i} - \theta_{i-1}$.
 We  ignore the coupling to the layers $i+2$ and $i-2$.
 
 \subsection{Numerical solution}
   We integrate the Langevin equations describing these three phases numerically, see  \cite{petersen1994some}, and depict the time evolution of 
 the variances $V_{w}(t) \equiv \langle \sin^{2}\theta_{w}(t)\rangle - \langle \sin \theta_{w}(t)\rangle^{2}$ and $\Vs\left(t\right) \equiv \langle \sin^{2}\theta_{s}(t)\rangle - \langle \sin \theta_{s}(t)\rangle^{2}$ in Fig.~\ref{fig:timeevolution}.
  $V_{w}$ and $V_{s}$ are a measure of the inter-layer phase fluctuations, and equally for the current fluctuations, keeping in mind that the currents across the Josephson junctions are  $j_{i} \equiv 2 J_{i} \sin(\theta_{i})$ with $i \in \{w,s\}$. 
 We use the parameters of Table \ref{tab:parameters}, and a driving frequency of $\wm = 2 \pi \times 10.4$ Thz, i.e. near the strong plasmon mode. 
  We use a temperature of $T=0.2$ K.  For this toy model, we have to use a temperature that is of the order of $J_{w}$, or else any phase coherence of the weak junction is suppressed. As we demonstrate below for the full, bulk model, temperatures of the order of $J_{ab}$ still give phase coherence of the weak junctions. 
 The driving amplitude $A_{0}$ is set to zero for $t<5$ ps,  and  $A_{0} = 5.2$ meV after that.
 
 We compare $V_{w}(t)$ and $V_{s}(t)$
 to their equilibrium values
\bea
V_{w/s, th} &=&  \langle\sin^{2}(\theta_{w,s})\rangle_{eq} = \frac{T}{\Jws}\frac{I_{1}(\Jws/T)}{I_{0}(\Jws/T)}\label{Vthermal}.
\eea
These are obtained by taking the expectation value for the equilibrium ensemble $\rho_{eq} = \exp((J_{w,s}/T)\cos(\theta_{w,s}))$.
 $I_{0}(x)$ and $I_{1}(x)$ are the modified Bessel functions of the first kind.   For the weak junction, we typically have $T \gg J_{w}$. In this limit we have
\bea
V_{w, th} &\approx& \frac{1}{2} - \frac{1}{16}\frac{J_{w}^{2}}{T^{2}}.\label{thermal}
\eea
 as the first terms of a high temperature expansion.
  As we see in Fig.~\ref{fig:timeevolution}, both variances undergo a 
    transient phase, during which $\Vw$ is visibly reduced. After that, a steady state emerges, in which the time average of  $V_{w}$ is smaller than in equilibrium. It is this reduction of fluctuations that we are interested in, and which is further enhanced in the bulk system discussed below. 
      We show the reduction of $V_{w}$ and $V_{s}$ in percent of the equilibrium value, which for the weak junction is  $1$ to $2$ percent. To state that reduction in more physical terms, we use Eq. \ref{Vthermal} as a measure, we translate $V_{w}$ into an effective temperature $T_{eff}$.  This gives a 'temperature' reduction of $5\%$.  As we demonstrate below, the state that is created via driving is not a thermal state, but rather a non-equilibrium state. $T_{eff}$ is purely an alternative measure of $V_{w}$.

\begin{figure}
        \includegraphics[scale=0.8]{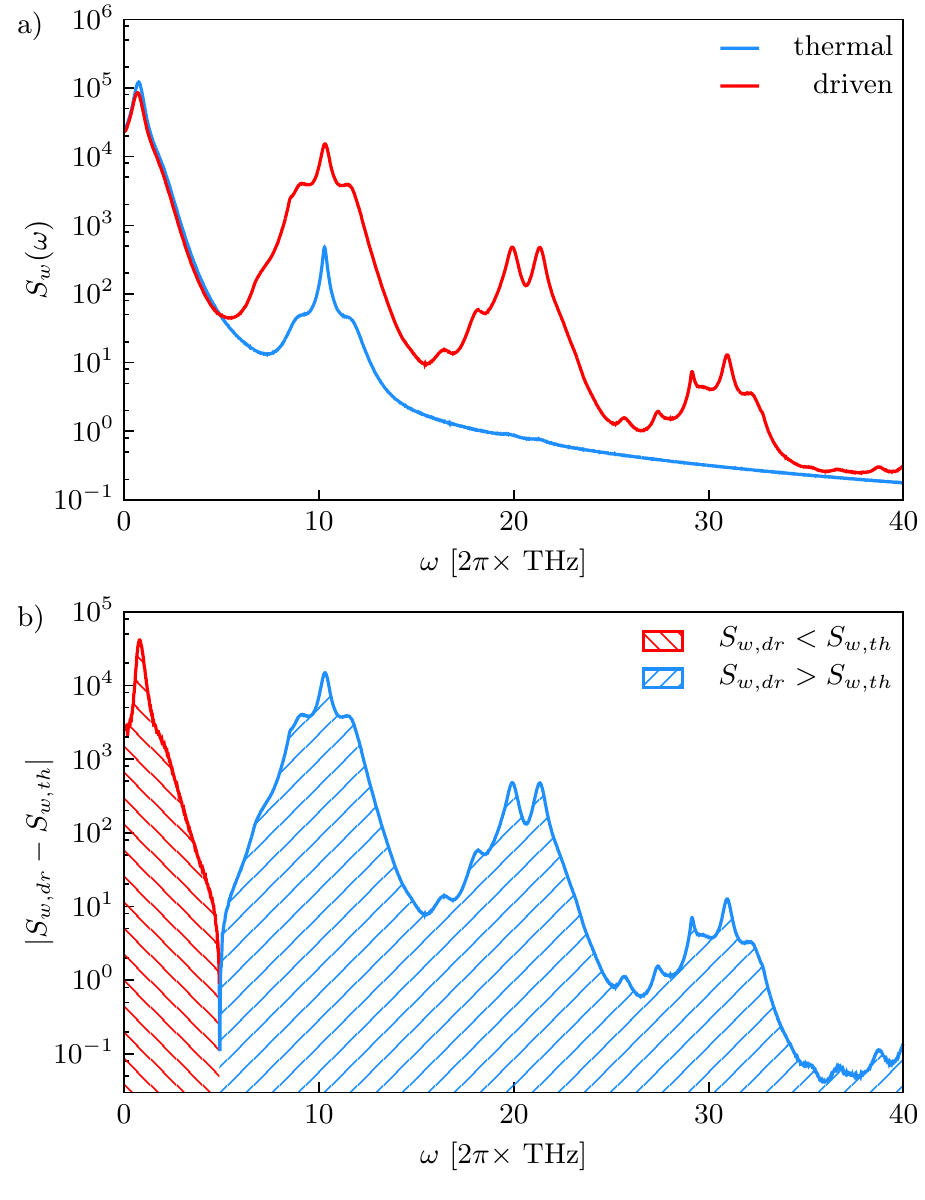}
    \caption{a) Power spectrum of the weak junction, $S_{w}(\omega)$, in arbitrary units on a logarithmic scale. The blue line is the thermal spectrum, the red line is the spectrum of the driven steady state.  b) Difference of the power spectrum of the driven state and the equilibrium spectrum. We find that the low-frequency fluctuations are reduced, whereas the fluctuations near the driving frequency are enhanced. Therefore, driving leads to an up-conversion of spectral weight.}
    \label{fig:toyPS}
\end{figure}

    In Fig.~\ref{fig:V_PS}  we show the time-averaged variances of the strong and weak junction, of the steady state, $\bar{V}_{w,s} \equiv \langle V_{w,s}(t)\rangle_{t}$, as a function of the driving frequency $\wm$. 
We observe a suppression of the fluctuations of the weak junction for a driving frequency near the high-energy plasmon frequency. We conclude that this suppression is not directly induced by the driving term in  Eq. \ref{eq:LD-hamiltonian-dr} operating on the weak junction, but by driving the high frequency mode near resonance, which in turn suppresses the fluctuations of the low-frequency mode. 
 The high frequency plasmon mode acts as an amplifier of the driving term acting on the weak junction. 
 
   In addition to the feature near $\omega_{s}$, there is a similar feature for driving frequencies near $\omega_{w}$. Here, $\bar{V}_{s}$ is unaffected, and the reduction is due to direct driving of the weak junctions. 
 We note however, that the driving frequencies in Refs. \cite{kaiser2012light, hu2013enhancement} are far away from the lower plasmon frequency, and focus on the phenomenon around $10$ THz. 
  The dependence of this result on the driving amplitude $A_{0}$ is discussed in Appendix \ref{toyV0}.

In Fig.~\ref{fig:toyPS}  we show the power spectrum of the currents in the weak junction for a driving frequency of $\omega_{m} = 2 \pi \times 10.4$ THz, where the effect is maximum, and for $A_{0} = 5.2$ meV. The power spectrum is defined as $S_{w}(\omega) \equiv \langle j_{w}(-\omega) j_{w}(\omega)\rangle - \langle j_{w}(-\omega)\rangle \langle j_{w}(\omega)\rangle$, where $j_{w}(\omega) = 1/\sqrt{T_{s}} \int dt' \exp(-i \omega t') j_{w}(t')$, with $T_{s}$ being the sampling time during the steady state. The power spectrum is therefore the Fourier transform of the two-time correlation function $\langle j_{w}(t_{1}) j_{w}(t_{2})\rangle$, with times $t_{1}$ and $t_{2}$ in the sampling time interval.

 We find that the fluctuations are reduced at low frequencies, which in equilibrium would correspond to a reduction of temperature. At the driving frequency and multiples of it, fluctuations are increased, resulting in a redistribution of phase fluctuations in frequency space. 
 
 This redistribution can be understood as follows. The integral over the power spectrum $\sum_{\omega} S_{w}(\omega)$ is essentially the time-averaged equal-time correlation of the current, i.e. $\bar{V}_{w}$. This quantity, however, is nearly saturated for $1/2$ at high temperatures, as can be seen from Eq. \ref{thermal}.  Therefore, the total area under the power spectrum has essentially reached its upper bound. Now, because the system is nonlinear, the high frequency modes near the driving frequency $\omega_{m}$ will be activated, and their weight in the power spectrum will increase. As a result the spectral weight in the low-frequency regime has to decrease. For this mechanism to occur, we therefore need two ingredients. Firstly, a non-linear system, for which the modes of different frequency interact, and secondly, high temperatures and a quantity whose fluctuations saturate at these temperatures. We indeed do not see this effect for a harmonic oscillator, or for low temperatures.

      \begin{figure*}
  \includegraphics[scale=0.85]{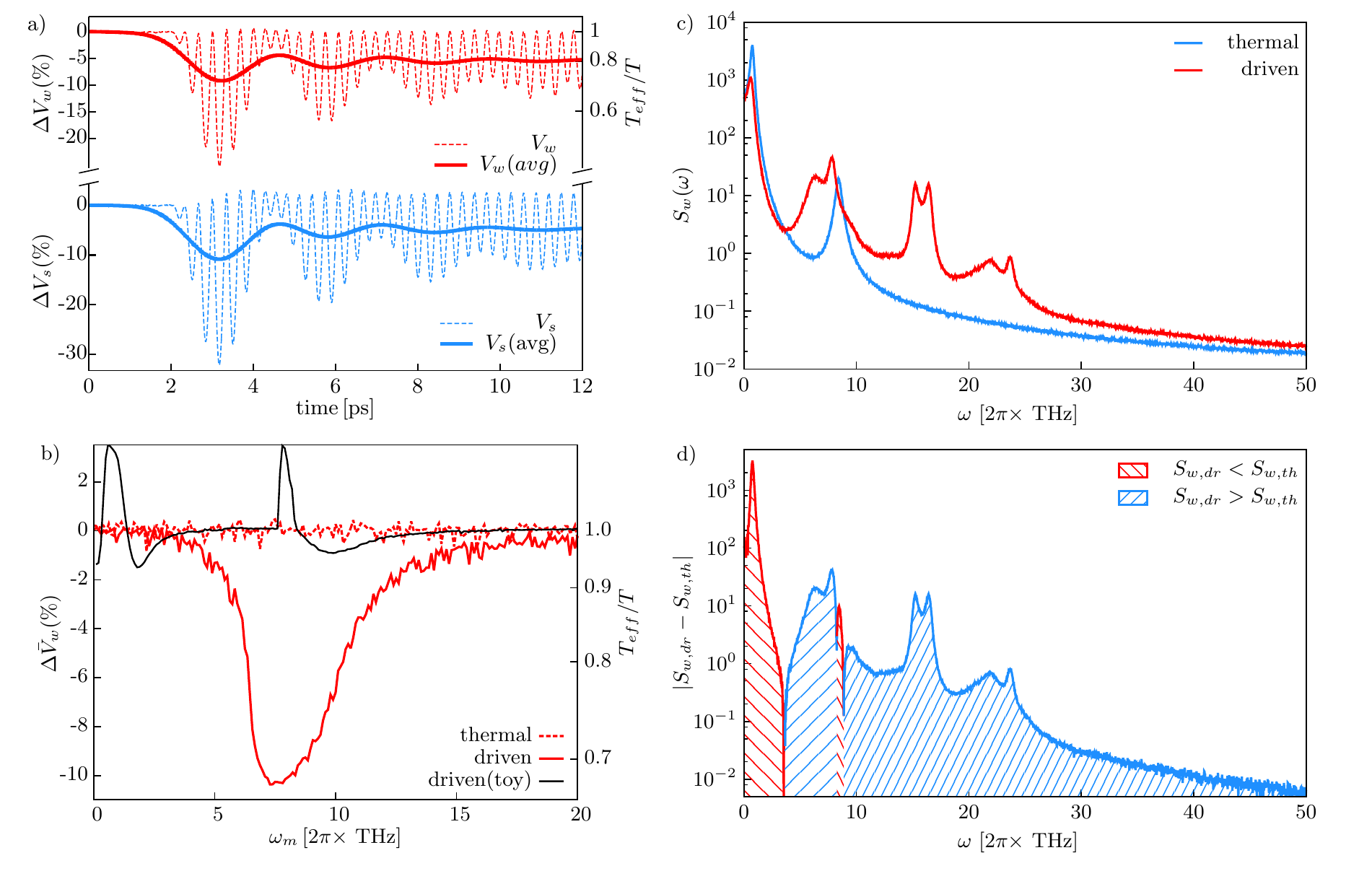}
    \caption{ (a) Time evolution of $V_{w}(t)$ and $V_{s}(t)$ of the bulk system  for a driving frequency of $\wm=2\pi\times10.4$ THz, a driving amplitude of $A_{0}=4.3$ meV, and a temperature of $T=100$ K.  The system is a $256\times 256\times 4$ lattice.
        (b) $\bar{V}_{w}$ of the bulk system, shown as a red, continuous line, plotted against the driving frequency, for a driving amplitude of $A_{0} = 4.3$ meV, for a $256\times 256\times 4$ system. For comparison, we show the non-driven value as a dashed line. 
 As a second comparison, we show $\bar{V}_{w}$ of the toy model as a black continuous line, with a temperature chosen such that the thermal magnitude of the bulk system is reproduced. We use a driving amplitude of $A_{0} = 4.3$ meV, which is near the optimal driving amplitude for the toy model.  We note that there is a large reduction of $\bar{V}_{w}$ near the resonance of the large plasmon frequency, and that the bulk system has a much stronger reduction than the toy model. Furthermore, the reduction due to direct driving of the weak junctions around $\omega_{w}$ is washed out due to the strong additional damping in the bulk. (c ) Power spectrum of the total current.  The system is a $128\times 128\times 4$ lattice at $T=100$ K, with a driving amplitude $A_{0}=4.3$ meV, and for a driving frequency of $\omega_{m} = 7.9$ THz, which is near the minimum of $\bar{V}_{w}$. We again see a reduction of the low-frequency fluctuations when the system is driven. (d) Difference of the power spectrum of the driven state and the equilibrium spectrum.}
    \label{fig:Bulk_frequency_scan}
\end{figure*}
   \begin{figure*}
    \includegraphics[scale=0.95]{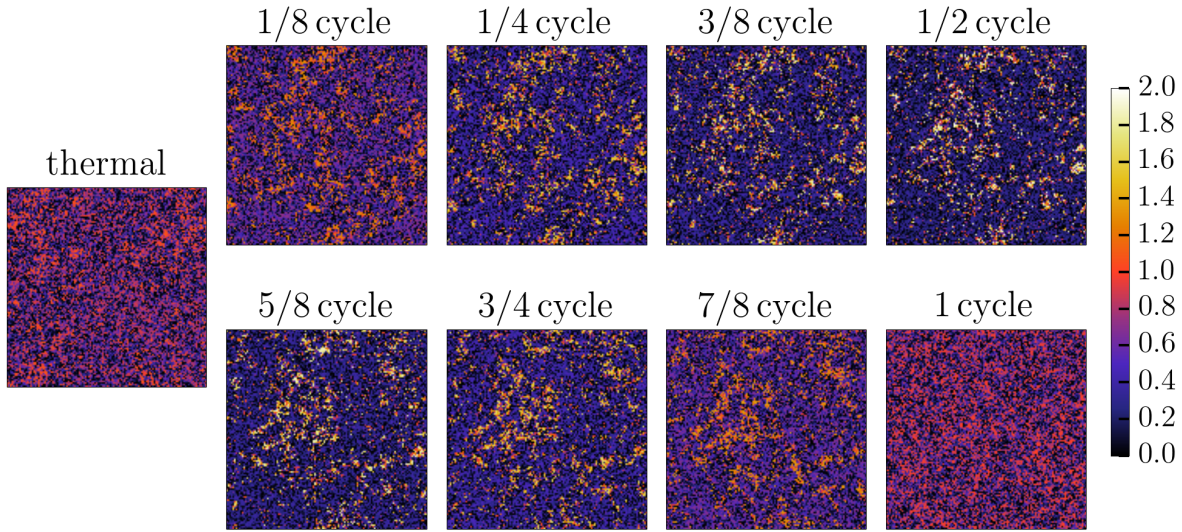}
    \caption{Current fluctuations in a plane of weak junctions, for a driving frequency of $\omega_{m} = 8.3$ THz, and for a driving amplitude of $A_{0} = 4.3$ meV. On the left, we see a thermal state, on the right, we show a cycle during the steady state of the driven system.}
    \label{fig:inplane}
\end{figure*}

\subsection{Analytical solution}
In this section we derive and discuss an analytical expression for the reduction of fluctuations, for  a limit of the two-oscillator toy model.
   The equations of motion for  the phase differences $\theta_{s}$ and $\theta_{w}$, that we discussed in the previous section, are
 \bea
    \ddot{\theta}_{w}  &=&   -\omega_{w}^2 \sin \theta_{w} + (\omega_{s}^2 \sin \theta_{s})/2 - \gamma\dot{\theta}_{w}+ \xi_{w}\nonumber\\
   && + 2 A_{0} (\wm \sin(\wm t)    - \gamma \cos(\wm t))   \label{eq:weakcoupledjj}\\
       \ddot{\theta}_{s}  &=&   -\omega_{s}^2 \sin \theta_{s} + (\omega_{w}^2 \sin \theta_{w})/2 - \gamma\dot{\theta}_{s}+ \xi_{s}\nonumber\\
   && - 2 A_{0} (\wm \sin(\wm t)    - \gamma \cos(\wm t)).   \label{eq:strongweakcoupledjj}
\eea 
 For the noise terms we assume $\langle \xi_{i} (t_{1}) \xi_{j}(t_{2})  \rangle = 4 E_{c} \gamma T  \delta_{ij} \delta(t_{1} - t_{2})$, with $i,j \in \{w,s\}$.
 We now consider the motion of the strong junction  as an external drive on the weak junction, see App. \ref{soa}. We combine this contribution and the external drive into an effective external driving term $F(t)=F_{0}\sin(\omega_{m} t)$. Furthermore, for calculational simplicity, we consider the overdamped limit:
\bea
\dot{\theta}_{w} = - \frac{\omega_{w}^{2}}{\gamma} \sin \theta_{w} + \frac{F(t)}{\gamma} + \frac{1}{\gamma} \xi(t),\label{effmodel}
\eea
 which is a driven, overdamped Josephson junction, coupled to a thermal bath, Ref. \cite{ojj}.
The corresponding Fokker-Planck equation, see e.g. Ref. \cite{van1992stochastic},  for $\theta_{w}$ is 
\bea
\partial_{t} \rho 
&= & \frac{2 T E_{c}}{\gamma} \partial_{\theta\theta} \rho + \frac{\omega_{w}^{2}}{\gamma} \partial_{\theta}(\sin(\theta) \rho)
 - \frac{F(t)}{\gamma} \partial_{\theta} \rho.\label{FP}
\eea
For notational simplicity, we have dropped the subscript 'w'.
 $\rho(\theta, t)$ is the time dependent probability distribution of $\theta$, defined on $(-\pi, \pi] $.   As described in App. \ref{FPderiv}, we choose a simple ansatz       $\rho = \exp(f(\theta,t))$, with 
\bea
f(\theta, t) &=& \frac{J_{w}}{T} \Big((1 + a_{c}(t)) \cos(\theta) +a_{s}(t) \sin(\theta) \Big)\label{fansatz}
\eea
 with two functions $a_{c}(t)$ and $a_{s}(t)$, which solve a set of of linear differential equations. 
  As demonstrated in Fig. \ref{fig:so} a), this solution captures both the transient and the steady state behavior.
  For the transient time scale we obtain $t_{tr} = \gamma/(2 T E_{c})$, which is indeed consistent with the numerical results. 
  For $\bar{V}_{w}$ we find 
\bea
\bar{V}_{w}  &\approx& V_{w, th} -\frac{1}{32}  \frac{J_{w}^{2}}{T^{2}} \frac{F_{0}^{2}}{\gamma^{2} \omega_{m}^{2}}\label{Vwapprox1}.
\eea
which indeed shows the reduction of fluctuations, compared to  $V_{w, th}$,  Eq. \ref{thermal}.
 In App. \ref{hiTexpand} we give a systematic high temperature expansion to fourth order in $F_{0}$ and find
\bea
\bar{V}_{w}  &\approx& V_{w, th} -\frac{1}{32}  \frac{J_{w}^{2}}{T^{2}} \frac{F_{0}^{2}}{\gamma^{2} \omega_{m}^{2}} +\frac{21}{512}  \frac{J_{w}^{2}}{T^{2}} \frac{F_{0}^{4}}{\gamma^{4} \omega_{m}^{4}}\label{VwF0},
\eea
  again written for $\gamma\omega_{m}\gg T E_{c}$. 
 Thus, in comparison to the equilibrium expression, the variance is first reduced, reaches a minimum at $F_{0}^{2} = (8/21) \gamma^{2} \omega_{m}^{2}$, and then increases again. In Fig. \ref{fig:so} b) we show Eq. \ref{VwF0} in comparison to the numerical solution of the single oscillator model, Eq. \ref{effmodel}, and the equilibrium value. We find that Eq. \ref{VwF0} captures the numerical result well, and that the suppression of fluctuations is even stronger than the analytical estimate. In App. \ref{hiTexpand} we find that a typical reduction of $T_{eff}$ for optimal driving  is between $5$ to $10\%$.

\section{Bulk system}\label{Sectbulk}
We next consider a more realistic model involving a stack of bi-layers, described by Eqs. \ref{eq:classical-EoM} and \ref{deltanLE}.
 We use either a lattice with $128\times 128$  or $256\times 256$ sites in the ab-plane, and $4$ in the c-direction. 
  We define $V_{w (s)}(t) \equiv (1/N_{w (s)}) \sum_{<ij>_{w(s)}}\langle \sin^{2}\theta_{ij}(t)\rangle - \langle \sin \theta_{ij}(t)\rangle^{2}$, where the sum is over all weak (strong) junctions, $N_{w (s)}$ is the number of weak (strong) junctions, and $\theta_{ij}$ is the phase difference between sites $i$ and $j$.
  The time evolution of $V_{w}$ and $V_{s}$ is shown in Fig. \ref{fig:Bulk_frequency_scan} a).
    In Fig. \ref{fig:Bulk_frequency_scan} b) we show the time average of the steady state $\bar{V}_{w,s} \equiv \langle V_{w,s}(t)\rangle_{t}$ as a function of the driving frequency. 
  The behavior that emerges from the extended model is qualitatively similar to the one described by the single bilayer model, see Figs. \ref{fig:timeevolution} and \ref{fig:V_PS}. However, the magnitude of the reduction of fluctuations is strongly enhanced.
   We note that the temperature of this example is significantly higher, while the magnitude of $V_{w}$ is comparable to the toy model examples. 
  This is due to the energy scale $J_{ab}$ which is indeed the main effective tunneling scale that the relative phase between two layers experiences.

 To derive an effective single oscillator model for the bulk model, we consider two neighboring pairs of sites, with the phases $\theta_{z,i}$, $\theta_{z,i+1}$, $\theta_{z+1,i}$ and $\theta_{z+1, i+1}$, where $z$ is the layer index, and $i$ is the site index in the plane. 
  The layers $z$ and $z+1$ are connected by  weak junctions. We go to the basis of phase differences across the weak junctions, $\theta_{w,j} \equiv \theta_{z+1, j} - \theta_{z,j}$ with $j \in\{ i, i+1 \}$, and the total phases, $\Theta_{w,j} \equiv (\theta_{z+1, j} + \theta_{z,j})/2$ with $j \in\{ i, i+1 \}$. The largest term in the Hamiltonian is the one that couples the phases in the planes.
   We therefore consider $H \approx -J_{ab}(\cos(\theta_{z+1, j+1} - \theta_{z+1,j}) + \cos(\theta_{z, j+1} - \theta_{z,j}))$.
 We write this expression in terms of $\theta_{w,j}$  and $\Theta_{w,j}$, and average out the fields $\Theta_{w,j}$, resulting in a temperature dependent prefactor. The remaining term is proportional to $\cos((\theta_{w, i} - \theta_{w,i+1})/2)$.
  We approximate this term via a mean-field decomposition  $\cos((\theta_{w, i} - \theta_{w,i+1})/2) \approx \cos((\theta_{w, i})/2) \langle  \cos((\theta_{w, i+1})/2)\rangle +\sin((\theta_{w, i})/2) \langle  \sin((\theta_{w, i+1})/2)\rangle$.
   The $\cos$ term is the effective non-linear oscillator contribution, whereas the $\sin$ term is an effective driving term.
    After rescaling $\theta_{w,i}/2 \rightarrow \theta_{w,i}$, we therefore end up with the same effective model as in Eq. \ref{effmodel}, where now both $J_{w} = J_{w, eff}(T)$ and $F_{0}$ are effective, temperature dependent parameters.
     However, the variance of the phase fluctuations is now 
   \bea
     V_{w, 2}(t) & \equiv & \langle \sin^{2}(2\theta_{w}(t))\rangle - \langle \sin(2 \theta_{w}(t))\rangle^{2}\label{Vw2}
    \eea
 because  of the field rescaling by $1/2$.
  In equilibrium it is
  \bea
  V_{w, 2} &=&  \frac{4 J_{w} T I_{1}(J_{w}/T) -12  T^{2}I_{2}(J_{w}/T)}{J_{w}^{2}I_{0}(J_{w}/T)}.\label{effmodelV}
  \eea
  We use this expression for $T_{eff}$ in Fig. \ref{fig:Bulk_frequency_scan} a). 
  For large temperatures this  approaches $V_{w} \approx \frac{1}{2} - \frac{1}{768}\frac{J_{w}^{4}}{T^{4}}$. 
    We can now use the solution of the Fokker-Planck equation for a single junction.
   As shown in App. \ref{FPderiv}, the time averaged value of $V_{w, 2}$ in  the driven steady state is
  \bea
  \bar{V}_{w, 2} &\approx& \frac{1}{2} - \frac{J_{w}^{4}}{768 T^{4}}  - \frac{19 J_{w}^{4}}{768 T^{4}}  \frac{F_{0}^{2}}{\gamma^{2} \omega_{m}^{2} + 4 T^{2} E_{c}^{2}}
  \eea
   which again shows a reduction due to driving.
We can also use the high temperature expansion of the FP equation, described in App. \ref{hiTexpand}, and calculate $\bar{V}_{w, 2}$. This indeed captures the magnitude of the reduction of phase fluctuations in the bulk, as shown in Fig. \ref{fig:BulkV0}.

In Fig.~\ref{fig:Bulk_frequency_scan} c) and d) we show the power spectrum $S(\omega)$ of the currents across a layer of weak junctions  
 $j_{w, tot}(t) \equiv  \sum_{<ij>,w} 2 J_{w} \sin\theta_{ij}$, i.e. $S(\omega) = \langle j_{w,tot}(-\omega) j_{w,tot}(\omega) \rangle$. Again we see that the fluctuations of  the low-frequency modes are reduced due to driving, similar to the toy model. 
   \begin{figure}
    \includegraphics[scale=0.9]{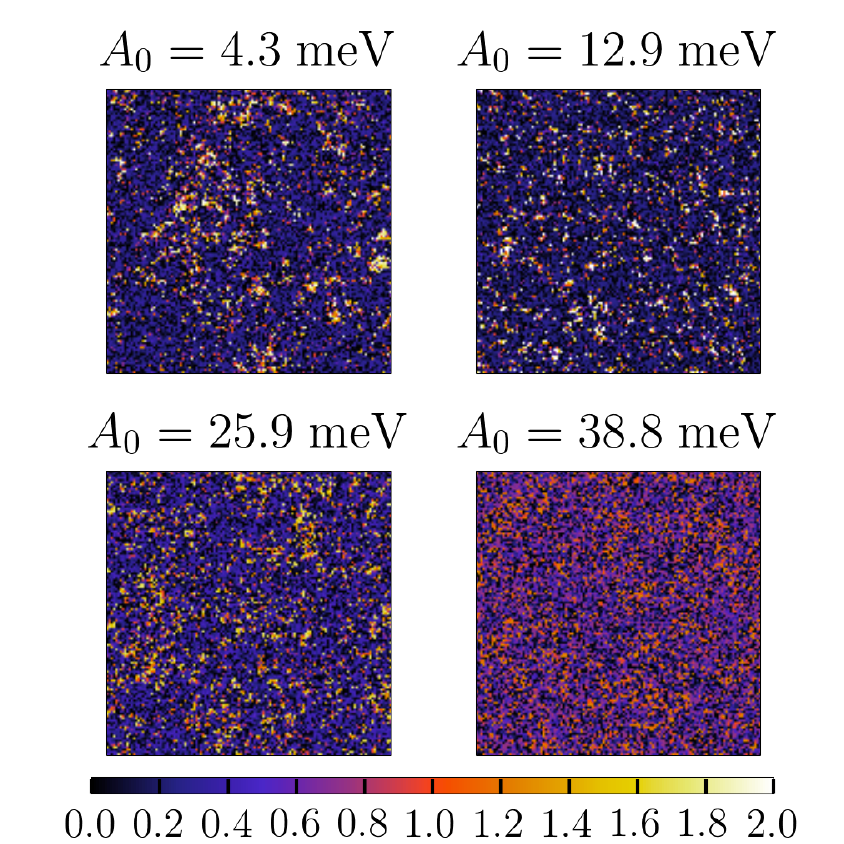}
    \caption{Current fluctuations in the steady state at half a cycle, for three values of $A_{0}$. The first panel is the same as in Fig. \ref{fig:inplane} at half a cycle.  $A_{0} = 12.9$ meV is near the optimal driving. At $A_{0} = 38.8$ meV, the magnitude of $\bar{V}_{w}$ has reached approximately its equilibrium value again, as can be seen from Fig. \ref{fig:BulkV0}.}
    \label{fig:inplaneA0}
\end{figure}

\section{In-plane dynamics}\label{SectInPlane}
Finally, we study the in-plane behavior of the driven bulk system. In Fig. \ref{fig:inplane} we show the fluctuations of the current
 $(j_{w}(\br, t) - \bar{j}_{w}(t))^{2}$, normalized by $1/(2 J_{w})^{2}$, for a single realization of the stochastic evolution of the system.
  We find that the system undergoes  periodic breathing  during a cycle. Furthermore, there are large regions, in which the fluctuations are suppressed, with smaller regions interspersed, in which the fluctuations are enhanced. 
  
  As described in the previous section, the magnitude of the inter-layer coherence $\bar{V}_{w}$ is first suppressed, as a function of the driving amplitude $A_{0}$, then reaches a minimum, before increasing again, see Fig. \ref{fig:BulkV0}. To illustrate how the in-plane current fluctuations are affected by this, we depict  $(j_{w}(\br, t) - \bar{j}_{w}(t))^{2}$ of single realizations, for increasing driving amplitude in Fig. \ref{fig:inplaneA0}. As is clearly visible, near the optimal driving amplitude the fluctuations are strongly suppressed, interspersed with small regions of increased fluctuations.

\begin{figure}
    \includegraphics[scale=0.8]{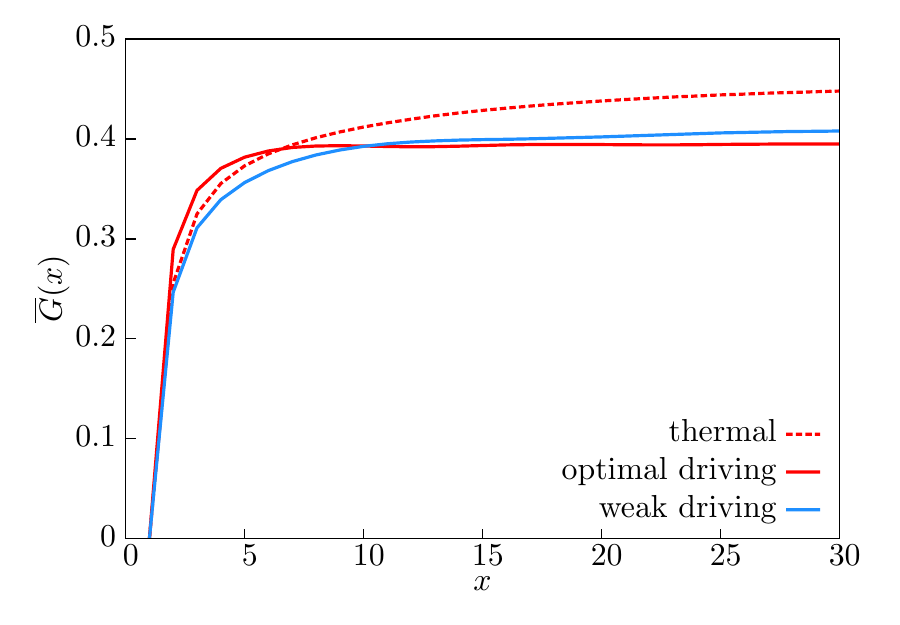}
    \caption{We show the equilibrium  in-plane  correlation function $G(\br)$ and the time averaged correlation function $\bar{G}(\br)$ of the driven state, for two driving amplitudes. Weak driving corresponds to $A_{0} = 4.3$ meV and optimal driving to $A_{0} = 12.9$ meV.}
    \label{fig:inplanecorr}
\end{figure}

  To study the in-plane behavior that was exemplified in  Figs. \ref{fig:inplane}  and \ref{fig:inplaneA0} quantitatively, 
 we  investigate how the current correlations between different sites within each plane is affected by driving.
 We define the current correlation function
 \bea
G(\br, t) &\equiv \sum_{\br_{0}} \frac{\langle (\sin(\theta_{w}(\br_{0}, t)) - \sin(\theta_{w}(\br_{0}+\br, t)))^{2}\rangle}{2N}.\label{def_ccf}
 \eea
 $\theta_{w}(\br, t)$ refers to the phase difference across a weak junction at the two dimensional site location $\br = (x,y)$ and at time $t$. The summation is over a single plane, with a number of sites $N$. In App. \ref{inplanebulk}, in particular in Fig. \ref{fig:currentcorr} we show the time evolution of this correlation function. 
  Based on the time evolution of $G(\br, t)$ we define the time average in the steady state $\bar{G}(\br) \equiv \langle G(\br, t)\rangle_{t}$.
  We depict this quantity in Fig. \ref{fig:inplanecorr}, for two values of the driving amplitude, in comparison to the equilibrium correlation function.
   We find that the current fluctuations are  visibly reduced  due to the driving, in particular on long scales. This would - in equilibrium - correspond to a reduced temperature. However,  we find that this asymptotic value is reached on a shorter scale, which indicates that the correlation length of the driven state is shorter. This is particularly visible for the driving amplitude near the optimum. 
   In equilibrium, the reduced correlation length would correspond to a higher temperature.   
  This again demonstrates that the resulting driven state is a non-equilibrium state, which cannot be captured by a single temperature on all scales. 
    This observation is consistent with the redistribution of phase fluctuations visible in the power spectra. The long-range modes behave as if the temperature has been reduced, whereas on short scales the system appears to be heated up.

The above suggests a possible physical frame for periodically driven bi-layer cuprates, which, in the limit of pre-existing pairs and of superconductivity being destroyed at the weakly coupled interbilayer junction by thermal phase fluctuations, may explain how important elements of the superconducting phase may persist or be re-established above $T_{c}$. Note that the response of the in-plane condensate is important in more than one respect. Firstly, it is possible that the stabilization of the long-range phase coherence may provide further stabilization for superconductivity at low frequency at the expense of enhanced in-plane fluctuations on short ranges and hence at higher frequency scales. Furthermore, although this is not studied here, the interaction of this driven phase with other coexisting or competing in-plane charge and spin orders, Refs. \cite{tranquada1995, ghiringhelli2013, sachdev2013},  may provide additional elements and prospects for dynamical stabilization in this class of compounds.

\section{Conclusions}\label{SectCon}
In summary, we have demonstrated that a reduction of thermal phase fluctuations in a layered superconductor can be achieved via external driving. We have developed an extended, anisotropic XY model to describe the dynamics of the pairing field, and a toy model that captures this effect.
 To give an example for the magnitude of the reduction in the bulk system, we again estimate the temperature $T_{eff}$ of the equilibrium system that gives the same variance $V_{w}$ as the driven system. We consider the data shown in Fig. \ref{fig:BulkV0}. We use Eq. \ref{effmodelV} to determine $J_{w, eff}$ from the equilibrium value of $V_{w} \approx 0.455$, which gives $J_{w, eff} \approx 341$ K. The reduction of $V_{w}$ to $\approx 0.385$ for optimal driving would correspond to an equilibrium temperature of $T_{eff} \approx 60$ K, compared to the equilibrium temperature $T = 100$ K.  This demonstrates  the remarkable reduction of fluctuations that is possible with this mechanism. 
With regards to the experiments reported in Refs. \cite{kaiser2012light, hu2013enhancement}, we note that the driving frequency was approximately $\approx 1.5$ that of the plasmon of the strong junctions. However, as is visible in Fig. \ref{fig:Bulk_frequency_scan} b), the response of the weak junction occurs in a broad frequency range above it, because of the amplifying effect of strong junction. Thus, the mechanism proposed here can be a possible explanation and contributing factor for the observations of Refs. \cite{kaiser2012light, hu2013enhancement}. 
 As for future experiments,  we have demonstrated that the suppression of phase fluctuations of the weak layers is most effective if the driving frequency is near the plasmon of the strong junction layer. We thus propose to use a material with an optical phonon mode near that plasmon frequency.

\begin{acknowledgments}
We acknowledge discussions with Alexander Lichtenstein, Maria Valentyuk, and Assa Auerbach.
 We acknowledge support from the Deutsche Forschungsgemeinschaft through the SFB 925 and the Hamburg Centre for Ultrafast Imaging, and from the Landesexzellenzinitiative Hamburg, supported by the Joachim Herz Stiftung. B.Z. acknowledges support from the China Scholarship Council, under scholarship No. 2012 0614 0012. 

\end{acknowledgments}


\appendix

\section{Dependence on the driving strength}\label{toyV0}
\begin{figure}
   \includegraphics[scale=0.9]{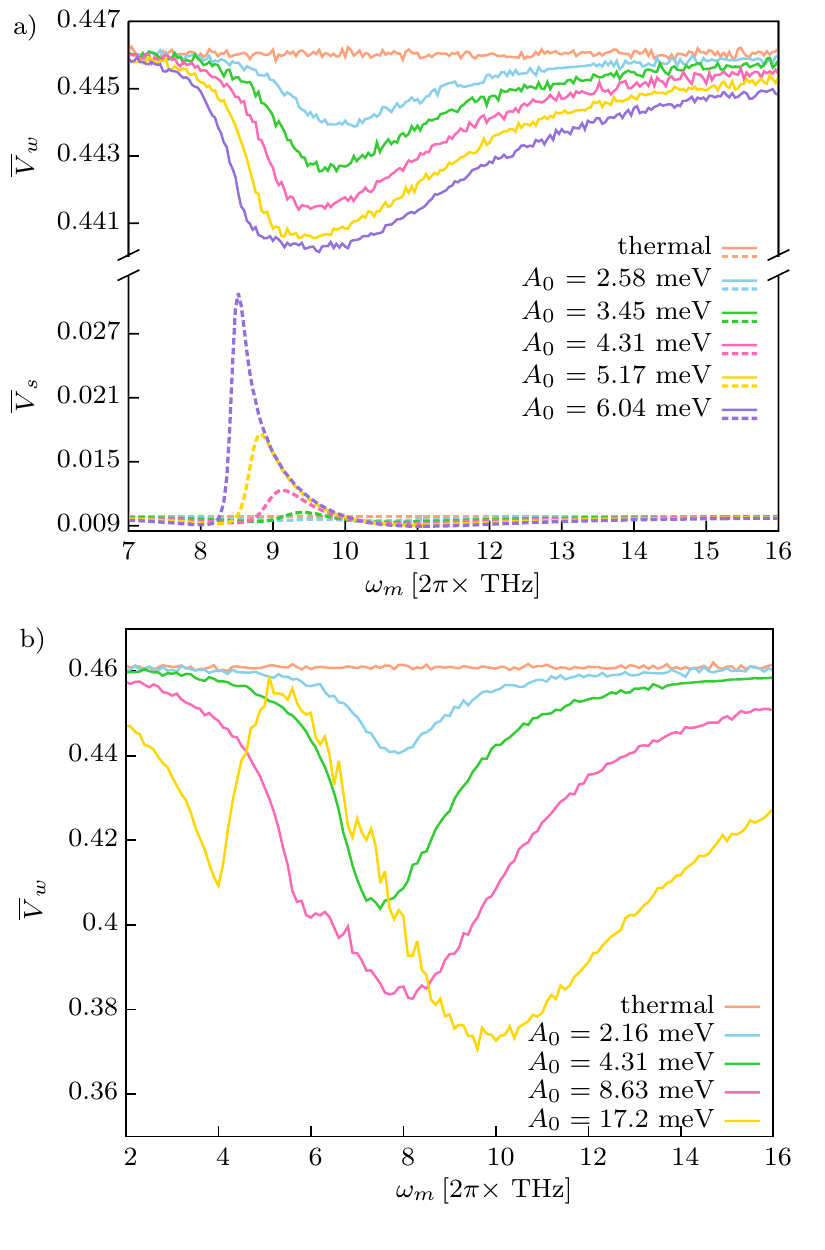} 
    \caption{(a) Time averaged variances $\bar{V}_{w}$ and $\bar{V}_{s}$ of the steady state of the toy model for increasing driving strength $A_{0}$. (b) The time averaged variance $\bar{V}_{w}$ for the bulk system, as a function of the driving frequency, for several values of $A_{0}$.}
    \label{fig:toyV0}
\end{figure}

To illustrate the dependence of $\bar{V}_{w}$ and $\bar{V}_{s}$ on the magnitude of the driving term $A_{0}$, 
 we  show frequency scans  for different values of $A_{0}$ in Fig. \ref{fig:toyV0} a), for the toy model. 
For computational simplicity, we choose the overdamped limit, with $\gamma = 2.1$ THz. We also choose $J_{w} = 0.25$ K, $J_{s} = 25$ K, $U = 5000$ K, and $T = 0.25$ K.  
 We observe that the reduction of $\bar{V}_{w}$ occurs over a large frequency range around the resonance frequency of the strong junction.
  As $A_{0}$ is increased, the response of the strong junction increases in magnitude. For small $A_{0}$, the amplitude of $\langle \sin(\theta_{s}(t))\rangle$ is small, and the  response is that of a driven harmonic oscillator. As $A_{0}$ is increased the non-linearity of the oscillator skews the response of the oscillator, as visible in $\bar{V}_{s}$ in Fig. \ref{fig:toyV0}. This response can be understood by expanding $\sin(\theta_{s})$ in the equation of motion, Eq. \ref{eq:strongweakcoupledjj},  to third order and using the solution of the driven Duffing oscillator. 
  The response of the weak junction shows a reduction of the fluctuations, which increases for increasing driving. As the driving is increased further, this effect is reverted and $\bar{V}_{w}$ is increased.

In Fig. \ref{fig:toyV0} b) we show the analogous frequency scans for the bulk system. We use the same parameters as above, with a temperature of $T = 200$ K. We again see a minimum of $\bar{V}_{w}$ if the system is driven near the resonance of the high energy plasmon. However, as the driving amplitude $A_{0}$ is increased, this tendency is reverted, and $\bar{V}_{w}$  increases again, similar to the behavior of the toy model.

\section{Driven, overdamped non-linear oscillator}\label{FPapp}
In this appendix, we elaborate on the analytical and numerical solutions of the driven, overdamped Josephson junction.
 In App. \ref{soa} we derive the effective single oscillator approximation, Eq. \ref{effmodel}, from the two oscillator model in Eqs. \ref{eq:weakcoupledjj} and \ref{eq:strongweakcoupledjj}. 
 In App. \ref{FPderiv} we discuss the ansatz in Eq. \ref{fansatz} for the Fokker-Planck equation, Eq.  \ref{FP}. 
 In App. \ref{hiTexpand} we discuss the high temperature expansion that gives the $\bar{V}_{w}$ estimate in Eq. \ref{VwF0}. 
 The high temperature expansion is also used for the comparison shown in Fig. \ref{fig:BulkV0}.

\subsection{Single oscillator approximation}\label{soa}
 In this section we elaborate on the single oscillator approximation, described by Eq. \ref{effmodel}.  
We approximate the strong junction as a driven harmonic oscillator, and ignore the coupling to the weak junction. In steady-state,  $\theta_{s}(t)$ is
\bea
\theta_{s}(t) &=& A \cos(\wm t) + B \sin(\wm t)\label{phis}
\eea
with 
\bea
A &=& 2 A_{0} \frac{\gamma \omega_{s}^{2}}{\left(\omega_{m}^{2} - \omega_{s}^{2}\right)^{2} + \gamma^{2} \omega_{m}^{2}}\\
B &=& 2 A_{0} \frac{\omega_{m} (\omega_{m}^{2}-\omega_{s}^{2}+\gamma^{2})}{(\omega_{m}^{2} - \omega_{s}^{2})^{2} + \gamma^{2} \omega_{m}^{2}}
\eea
The skewness of the response of the strong junction, which is also visible in Fig. \ref{fig:toyV0}, is due to the non-linearity of the oscillator. It can be understood by expanding the $\sin \theta_{s}$ to cubic oder, and using the Duffing oscillator solution.
 For the weak junction, we consider the overdamped limit, where we ignore the $\ddot{\theta}_{w}$ term:
\bea
\dot{\theta}_{w} = - \frac{\omega_{w}^{2}}{\gamma} \sin \theta_{w} + \frac{F(t)}{\gamma} + \frac{1}{\gamma} \xi\left(t\right),
\eea
 which is Eq. \ref{effmodel}. 
 We interpret $\theta_{s}(t)$ as an external driving term, given by Eq. \ref{phis}, and linearize $\sin(\theta_{s}(t))$.  The resulting driving term for the weak junctions is
 $F(t)$:
\bea
F(t) &=& F_{c} \cos (\wm t) + F_{s} \sin(\wm t) \label{eq:effextforc}
\eea
where $F_{c} = \omega_{s}^{2} A/2 - 2 \gamma A_{0}$ and $F_{s} =  \omega_{s}^{2} B/2 + 2 \wm A_{0}$. 
 We write this as $F(t) = F_{0} \sin(\wm t + \phi_{0})$, with $F_{0} = \sqrt{F_{c}^{2} + F_{s}^{2}}$ and $\phi_{0} = \arctan(F_{c}/F_{s})$. We shift the time axis $t \rightarrow t - \phi_{0}/\wm$, and the resulting driving term is $F(t) = F_{0} \sin(\wm t)$.

\begin{figure}
    \includegraphics[scale=0.8]{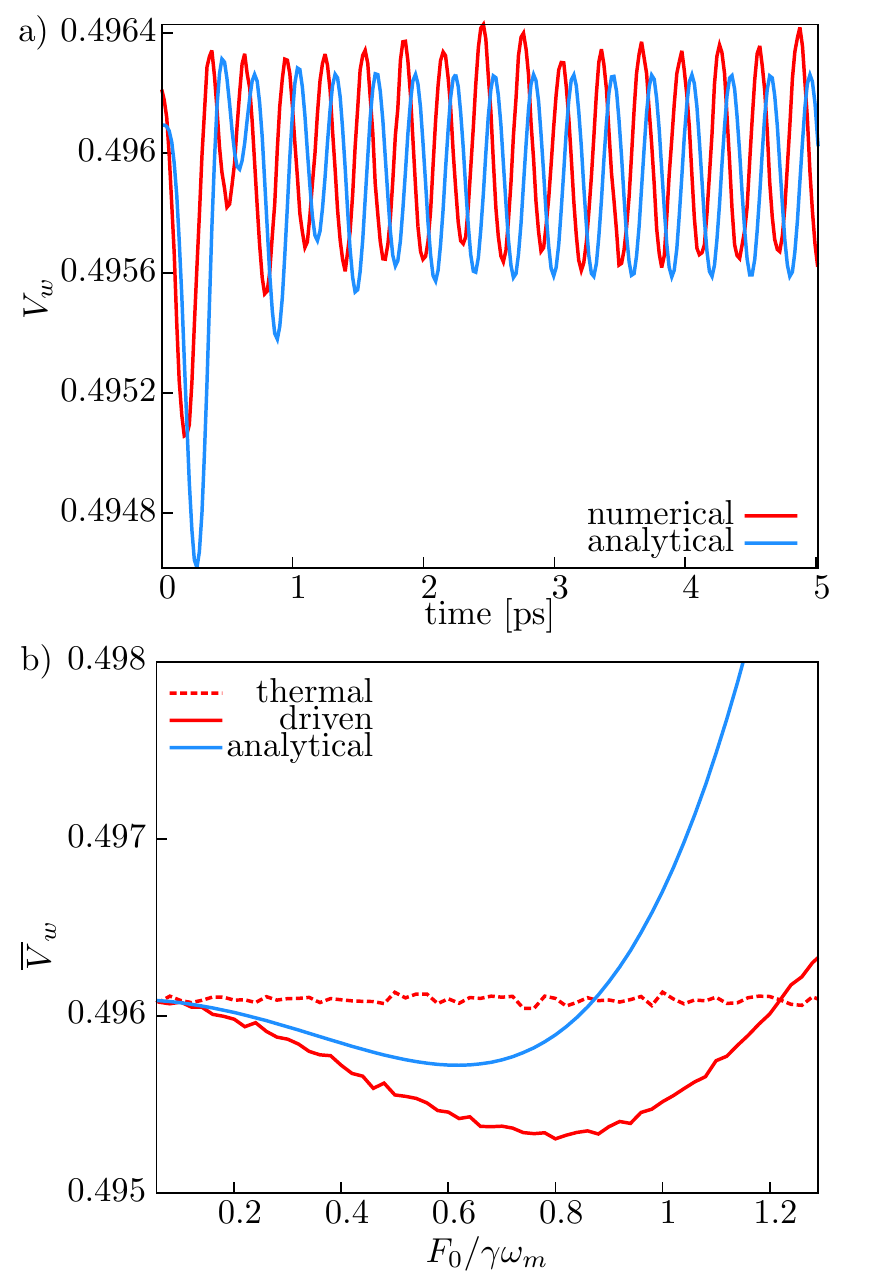} 
    \caption{a) Time evolution of $V_{w}$ for the effective, single oscillator model, Eq. \ref{effmodel}. The temperature is $T = 1$ K, the damping $\gamma = 2.1$ THz, the driving frequency $\omega_{m} =10.4$ THz, $U$ is $5000$ K, and $F_{0}/(\gamma \omega_{m}) =0.3$.
         The red line shows the numerical solution, the blue line the analytical solution in Eq. \ref{Vwapprox}.  b) Time averaged $\bar{V}_{w}$ for the single oscillator model, as a function of the driving amplitude. The red lines are numerical results, the blue line is the analytical result in Eq. \ref{VwF0}.}
    \label{fig:so}
\end{figure}

 \begin{figure}
    \includegraphics[scale=0.9]{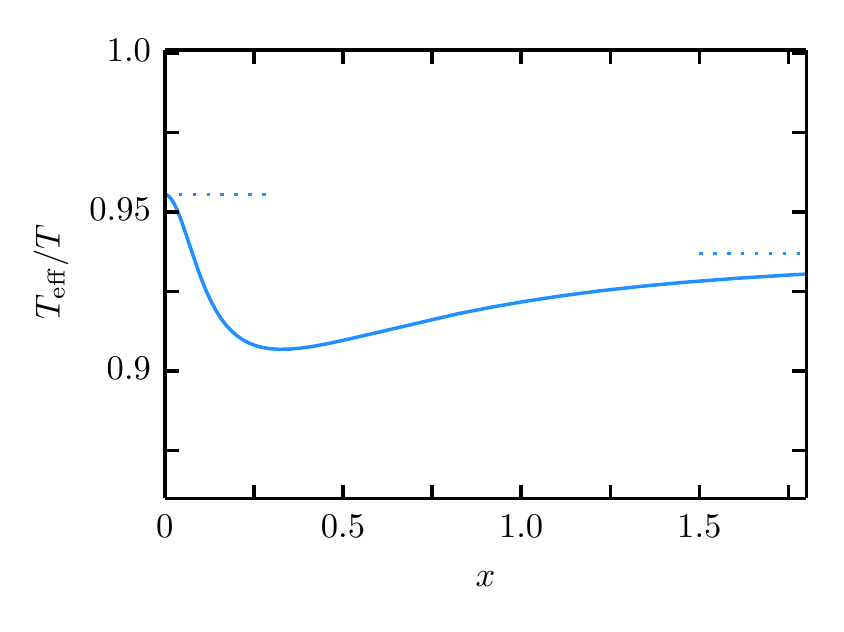} 
    \caption{Effective temperature as a function of $x$. The dashed lines indicate the asymptotic values for $x \rightarrow 0$ and $x \rightarrow \infty$.}
    \label{fig:Teff}
\end{figure}

 \subsection{Ansatz for the Fokker-Planck equation}\label{FPderiv}
In this section we discuss the approximate, analytical solution of the Fokker-Planck equation, based on the ansatz in Eq. \ref{fansatz}. 
 In equilibrium and without driving, the Fokker-Planck equation, Eq. \ref{FP},  is solved by
\bea
\rho_{0} &=& \exp\Big(\frac{J_{w}}{T} \cos(\theta)\Big)
\eea
To solve the driven case, we write $\rho(t)$ as $\rho = \exp(f(\theta,t))$ and obtain
\bea
\partial_{t} f &=& \frac{2 E_{c}}{\gamma}\Big(T(\partial_{\theta \theta} f + (\partial_{\theta} f)^{2}) +J (\cos \theta + \sin \theta \partial_{\theta} f)\Big)\nonumber\\
&& - F(t) \partial_{\theta} f/\gamma\label{feq}
\eea
$f$ has to be periodic in $\theta$, and can therefore be expanded in a Fourier series $f = \sum_{n} \exp(i n \theta) f_{n}$.
  For the high temperature and weak driving regime, we limit this expansion to the first harmonic, $n=1$. We consider 
\bea
f(\theta, t) &=& \frac{J_{w}}{T} \Big((1 + a_{c}(t)) \cos(\theta) +a_{s}(t) \sin(\theta) \Big)
\eea
With this, we ignore the higher harmonic terms in Eq. \ref{feq}, in particular the terms $\sim (\partial_{\theta} f)^{2}$ and $\sim \sin \theta \partial_{\theta} f$. 
 The resulting equations of motion for $a_{s}$ and $a_{c}$ are 
\bea
\dot{a}_{s} &=& -\frac{2 T E_{c}}{\gamma} a_{s} + \frac{F_{0}}{\gamma} \sin(\omega_{m} t)\label{aseq}\\
\dot{a}_{c} &=& -\frac{2 T E_{c}}{\gamma} a_{c} - \frac{F_{0}}{\gamma} \sin(\omega_{m} t)  a_{s}\label{aceq}
\eea
where we linearized Eq. \ref{aseq}, with the assumption $a_{s},a_{c} \ll 1$.
With $a_{s,c}(0) = 0$, these are solved by
\bea
a_{s}(t) &=& C_{1} \exp(- t/t_{tr}) - C_{1}\cos(\omega_{m} t)\nonumber\\
&& + C_{2} \sin(\omega_{m}t)\\
a_{c}(t) &=& \exp(- t/t_{tr}) (C_{3} + C_{4} \cos(\omega_{m} t)) - C_{4}/2\nonumber\\
&& - C_{5} \cos(2\omega_{m} t) + C_{6} \sin(2 \omega_{m} t)
\eea
with
\bea
C_{1} &=& \frac{F_{0} \gamma \omega_{m}}{\gamma^{2} \omega_{m}^{2} + 4 T^{2} E_{c}^{2}}\\
C_{2} &=& \frac{2 F_{0} T E_{c}}{\gamma^{2} \omega_{m}^{2} + 4 T^{2} E_{c}^{2}}\\
C_{3} &=& -\frac{F_{0}^{2}}{4} \frac{1}{\gamma^{2} \omega_{m}^{2} + T^{2} E_{c}^{2}}\\
C_{4} &=& F_{0}^{2} \frac{1}{\gamma^{2} \omega_{m}^{2} + 4 T^{2} E_{c}^{2}}\\
C_{5} &=& \frac{F_{0}^{2}}{4} \frac{\gamma^{2} \omega_{m}^{2} -2 T^{2} E_{c}^{2}}{\gamma^{4} \omega_{m}^{4} + 5 \gamma^{2} \omega_{m}^{2} T^{2} E_{c}^{2}+ 4 T^{4} E_{c}^{4}}\\
C_{6} &=& \frac{3 F_{0}^{2}}{4} \frac{\gamma \omega_{m}  T E_{c}}{\gamma^{4} \omega_{m}^{4} + 5 \gamma^{2} \omega_{m}^{2} T^{2} E_{c}^{2}+ 4 T^{4} E_{c}^{4}}
\eea
$t_{tr}$ is the transient time scale, $t_{tr} = \gamma/(2 T E_{c})$. This time scale increases with increasing damping, as it is typically the case in the overdamped limit.
 We note that for small temperatures, in particular for $T E_{c}\ll \gamma \omega_{m}$, the driving amplitude $F_{0}$ has to be compared to 
  the energy scale $\gamma \omega_{m}$, while  for $T E_{c}\gg \gamma \omega_{m}$, it has to be compared to  $T E_{c}$.
   This can already be read off from, say, Eq. \ref{aseq}. Since the derivative $\dot{a}_{s}$ is approximately $\sim \omega_{m} a_{s}$ in the driven state, there are two homogeneous terms to counter the driving term. As a result,  $a_{s}$ will scale as $\sim F_{0}/(\gamma \omega_{m})$ or as 
  $\sim F_{0}/(T E_{c})$, depending on which is the dominant energy scale.

  \begin{figure}
    \includegraphics[scale=0.95]{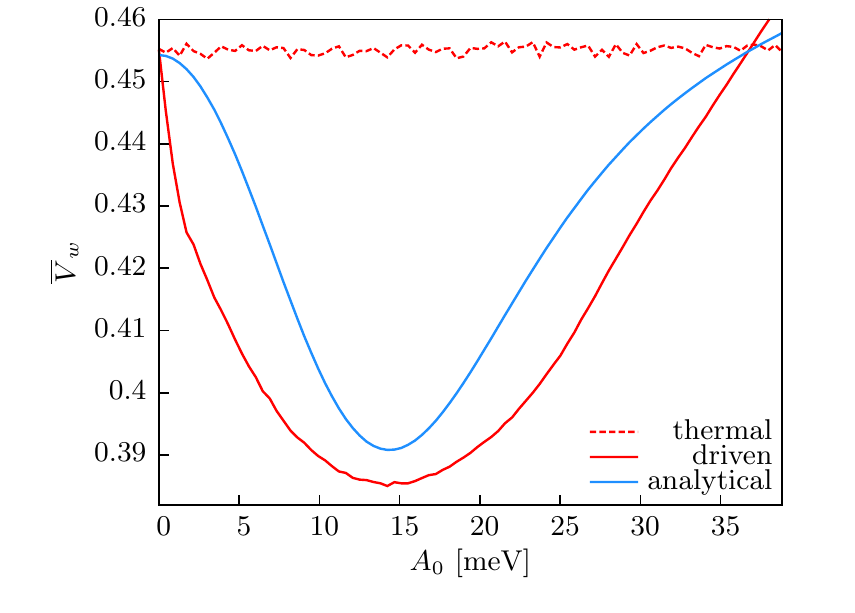}
    \caption{We show $\bar{V}_{w}$ as a function of the driving amplitude $A_{0}$, for the bulk system of $256 \times 256 \times 4$ sites, at $T=100$ K, at a driving frequency of $\omega_{m} = 8.3$ THz. Additionally, we show 
  the prediction of the effective model, calculated via the high temperature expansion of the FP equation described in App. \ref{hiTexpand}. The result for the bulk simulation is shown as a continuous, red line, with the non-driven case as a dashed, red line. The result for the effective single junction model is shown in blue.
     $J_{w, eff}$ has been chosen so that the equilibrium value of $V_{w}$ is reproduced. $F_{0}$ has been chosen to be proportional to $V_{0}$, with a proportionality coefficient such that the minima match up.    We indeed see that the large reduction of the phase fluctuations is approximately captured by the effective single oscillator model.}
    \label{fig:BulkV0}
\end{figure}

 $V_{w}(t)$ is given by  
 \bea
 V_{w}(t) &\approx& \frac{1}{2} - \frac{J_{w}^{2}}{16 T^{2}}(1 + 2 a_{c}(t) + 3 a_{s}^{2}(t))\label{Vwapprox}
 \eea
 within the high temperature expansion, and up to second order in $F_{0}$.  We note that $a_{s}$ scales as first order in $F_{0}$, and $a_{c}$ as second order.
 This high temperature expansion in is in analogy to Eq. \ref{thermal}. 
  In Fig. \ref{fig:so} a) we compare Eq. \ref{Vwapprox} to the numerical solution. We find that both the transient behavior and the steady state are captured by the analytical expression.
  In the steady state, we have
 \bea
 \langle a^{2}_{s}\rangle &=& - \langle a_{c}\rangle = \frac{F_{0}^{2}}{2} \frac{1}{\gamma^{2} \omega_{m}^{2} + 4 T^{2} E_{c}^{2}}
 \eea
 Therefore, the time averaged value of $V_{w}$ in the steady state is
 \bea
 \bar{V}_{w} &\approx& \frac{1}{2} - \frac{J_{w}^{2}}{16 T^{2}}  - \frac{J_{w}^{2}}{32 T^{2}}  \frac{F_{0}^{2}}{\gamma^{2} \omega_{m}^{2} + 4 T^{2} E_{c}^{2}}
 \eea
  For $\gamma \omega_{m}\gg T E_{c}$ this reduces to Eq. \ref{Vwapprox1}.

    As described in the discussion of the bulk system, we can use a single oscillator for the bulk system as well. However, we have to consider
  the variance of  $\sin(2 \theta)$, ${V}_{w, 2}(t)$, defined in Eq. \ref{Vw2}, rather that $V_{w}$. This is given by
 \bea
 V_{w, 2}(t) &\approx& \frac{1}{2} - \frac{J_{w}^{2}}{768 T^{2}}(1 + 4 a_{c}(t) + 42 a_{s}^{2}(t))
 \eea
 The time averaged value for the steady state is 
 \bea
 \bar{V}_{w, 2} &\approx& \frac{1}{2} - \frac{J_{w}^{4}}{768 T^{4}}  - \frac{19 J_{w}^{4}}{768 T^{4}}  \frac{F_{0}^{2}}{\gamma^{2} \omega_{m}^{2} + 4 T^{2} E_{c}^{2}}
 \eea

 \subsection{High temperature expansion}\label{hiTexpand}
  As a more systematic approach, 
  we expand $\rho$  as $\rho\left(\phi, t\right) = \frac{1}{2\pi} \sum_{k, n} \exp\left(i k \phi + i n \wm t\right)\rho_{k, n}$. The FP equation for $\rho_{k, n}$ is
 \bea
i n \wm \rho_{k, n} &=& - \frac{2 T E_{c}}{\gamma} k^{2} \rho_{k, n} + \frac{\Jw E_{c}}{\gamma} k ( \rho_{k-1, n} -  \rho_{k+1, n})\nonumber\\
 &&  - \frac{F_{0}}{2\gamma} k (\rho_{k, n-1} -  \rho_{k, n+1})\label{FPmatrix}
 \eea
We reduce $\rho_{k,n}$ to a finite number of coefficients by taking into account only $k = -2, \dots, 2$ and $n=-2, \dots, 2$, for the analytical result for $\bar{V}_{w}$ shown below and in Eq. \ref{VwF0}, and $k = -8, \dots, 8$ and $n=-8, \dots, 8$ for the numerical solution depicted  in Fig. \ref{fig:BulkV0}.

We write out the case of $k = -2, \dots, 2$ and $n=-2, \dots, 2$. Extending the range of these coefficients can be easily done by analogy. 
 We represent the coefficients $\rho_{k,n}$, with $k=-2,\dots, 2$ and $n=-2,\dots, 2$, as a single column vector with the entries $\tilde{\rho} \equiv (\rho_{2,2}, \rho_{2,1}, \rho_{2,0}, \rho_{2,-1}, \rho_{2,-2}, \rho_{1,2}, \dots, \rho_{-2,-2})$. The Fokker Planck Eq. \ref{FPmatrix} is then
\bea
i\wm M_{0} \tilde{\rho} &=& - \frac{2 T E_{c}}{\gamma} M_{1} \tilde{\rho} + \frac{\Jw E_{c}}{\gamma} M_{2} \tilde{\rho}\nonumber\\
&& - \frac{F_{0}}{2 \gamma} M_{3} \tilde{\rho}
\eea
with
\bea
M_{0} &=& K_{0}  \otimes {\bf 1}_{5}\\
M_{1} &=& {\bf 1}_{5}\otimes K_{1}\\
M_{2} &=& {\bf 1}_{5}\otimes K_{2}\\
M_{3} &=& K_{3} \otimes K_{0}
\eea
and
\bea
K_{0} &=& \left( \begin{array}{ccccc}
2 & 0 & 0 & 0 &0\\
0 & 1 & 0 & 0 & 0 \\
0 & 0 & 0 & 0 & 0 \\
0 & 0 & 0 & -1 & 0 \\
0 & 0 & 0 & 0 & -2 
\end{array} \right)\\
K_{1} &=& \left( \begin{array}{ccccc}
4 & 0 & 0 & 0 &0\\
0 & 1 & 0 & 0 & 0 \\
0 & 0 & 0 & 0 & 0 \\
0 & 0 & 0 & 1 & 0 \\
0 & 0 & 0 & 0 & 4 
\end{array} \right)\\
K_{2} &=& \left( \begin{array}{ccccc}
0 & 2 & 0 & 0 &0\\
-1 & 0 & 1 & 0 & 0 \\
0 & 0 & 0 & 0 & 0 \\
0 & 0 & 1 & 0 & -1 \\
0 & 0 & 0 & 2 & 0 
\end{array} \right)\\
K_{3} &=& \left( \begin{array}{ccccc}
0 & 1 & 0 & 0 &0\\
-1 & 0 & 1 & 0 & 0 \\
0 & -1 & 0 & 1 & 0 \\
0 & 0 & -1 & 0 & 1 \\
0 & 0 & 0 & -1 & 0 
\end{array} \right)
\eea
We solve this set of linear equations, and evaluate $\Vw\left(t\right)$. We expand to second order in $1/T$ and to fourth order in $F_{0}$. This gives $\Vw \left(t\right)$, from which we calculate the time averaged value $\bar{V}_{w}$ to be
\bea
\bar{V}_{w} 
 &\approx& \frac{1}{2} -\frac{1}{16} \frac{J_{w}^{2}}{T^{2}}
 -\frac{1}{32}  \frac{J_{w}^{2}}{T^{2}} \frac{F_{0}^{2}}{\gamma^{2} \omega_{m}^{2}}  f_{2}(x)\nonumber\\
&& +\frac{3}{512}  \frac{J_{w}^{2}}{T^{2}} \frac{F_{0}^{4}}{\gamma^{4} \omega_{m}^{4}} f_{4}(x)\label{Vwbarfull}
 \eea
with
\bea
f_{2}(x) &=&  \frac{1 + 144 x^{2}}{1 + 68 x^{2} + 256 x^{4}}\\
f_{4}(x) &=& \frac{7 + 948 x^{2}+ 15648 x^{4} + 24832 x^{6}}{(1 + 64 x^{2}) (1 + 4 x^{2})^{2} (1 + 17 x^{2} + 16 x^{4})}
\eea
and  $x \equiv T E_{c}/(\gamma \wm)$. 
For $\gamma \wm \gg T E_{c}$, i.e. $x\ll 1$, the expression for $\bar{V}_{w}$ simplifies to 
\bea
\bar{V}_{w}  &\approx& \frac{1}{2} -\frac{1}{16} \frac{J_{w}^{2}}{T^{2}} -\frac{1}{32}  \frac{J_{w}^{2}}{T^{2}} \frac{F_{0}^{2}}{\gamma^{2} \omega_{m}^{2}}\\
&& +\frac{21}{512}  \frac{J_{w}^{2}}{T^{2}} \frac{F_{0}^{4}}{\gamma^{4} \omega_{m}^{4}},
\eea
 which is the same as  Eq. \ref{VwF0}. 
 We also note that the second order term in $F_{0}$ is the same as in Eq. \ref{Vwapprox1}. 
 This expression is minimized for $F_{0}^{2} = (8/21) \gamma^{2} \wm^{2}$. The resulting minimal value for $\bar{V}_{\text{w}}$ is $\bar{V}_{\text{w}} = 1/2 - (23/336)J_{w}^{2}/T^{2}$. 
 If we formally equate this to $1/2 - J_{w}^{2}/(16 T_{eff}^{2})$, purely as a measure of the reduction of the fluctuations, we obtain an effective temperature of $T_{eff}/T = \sqrt{21/23} \approx 0.96$. 

\begin{figure}
    \includegraphics[scale=0.27]{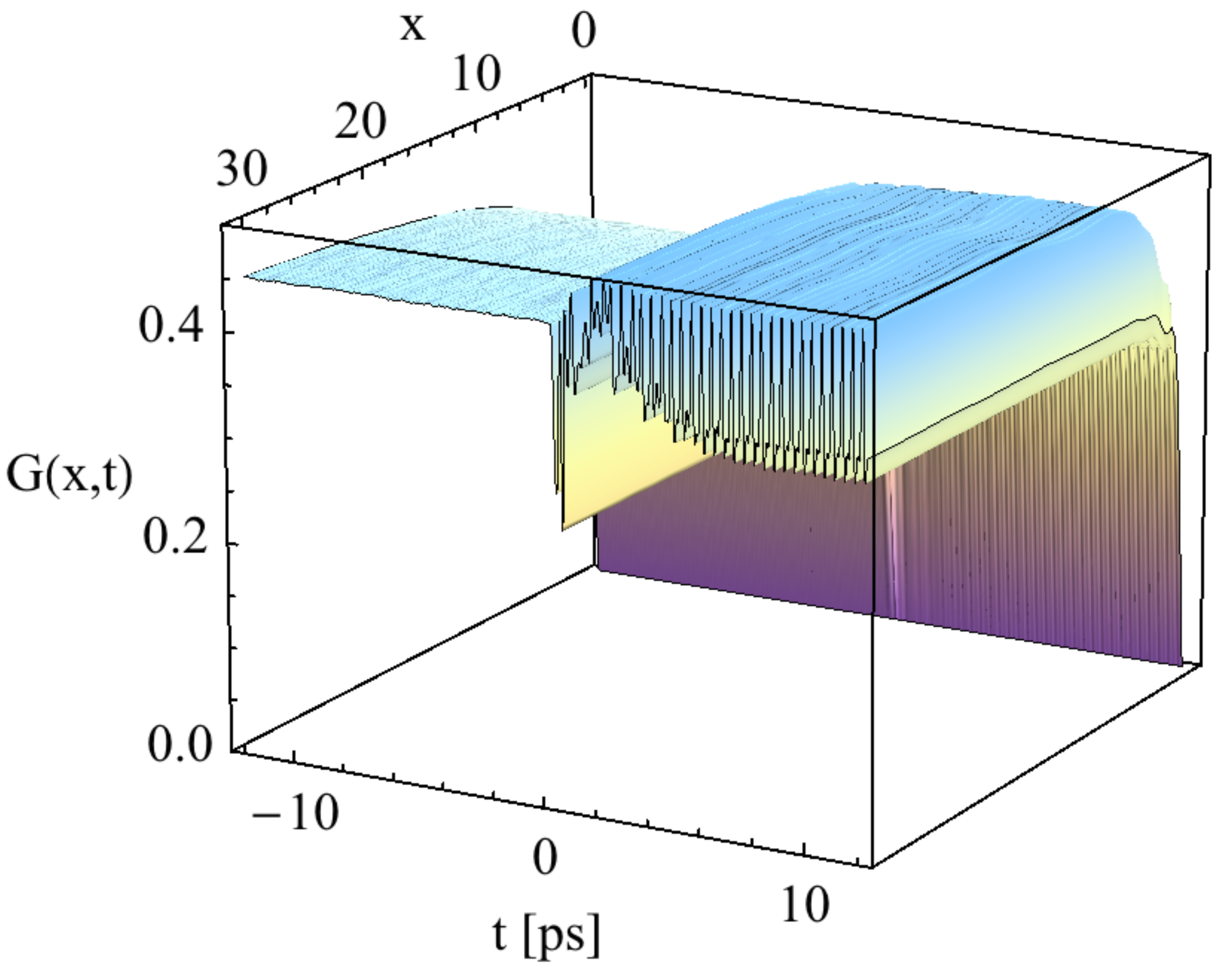} 
    \caption{Time evolution of the current correlation function, defined in Eq. \ref{def_ccf}.}
    \label{fig:currentcorr}
\end{figure}
 In the opposite limit of  $\gamma \wm \ll T E_{c}$ we have 
\bea
\bar{V}_{w}  &\approx& \frac{1}{2} -\frac{1}{16} \frac{J_{w}^{2}}{T^{2}} -\frac{9}{512}  \frac{J_{w}^{2}}{T^{2}} \frac{F_{0}^{2}}{T^{2} E_{c}^{2}}\\
&& +\frac{291}{32768}  \frac{J_{w}^{2}}{T^{2}} \frac{F_{0}^{4}}{T^{4} E_{c}^{4}}
\eea
This is minimized for $F_{0}^{2}/(T^{2} E_{c}^{2}) = (192/97)$. The minimal value of $\bar{V}_{w}$ is $\bar{V}_{w} = 1/2 - (221/3104) J_{w}^{2}/T^{2}$, and the effective temperature is $T_{eff}/T = \sqrt{194/221} \approx 0.94$. 

The full expression for $\bar{V}_{\text{w}}$ in Eq. \ref{Vwbarfull} is minimized for
\bea
F_{0}^{2} &=& \frac{8}{3} \frac{f_{2}(x)}{f_{4}(x)} \gamma^{2} \omega_{m}^{2}
\eea
For this value of $F_{0}$, $\bar{V}_{w}$ is
\bea
\bar{V}_{\text{w}} &\approx& \frac{1}{2} -\frac{1}{16} \frac{\Jw^{2}}{T^{2}} - \frac{1}{24} \frac{\Jw^{2}}{T^{2}} \frac{f_{2}(x)^{2}}{f_{4}(x)}
\eea
 We again formally equate this to $1/2 - J_{w}^{2}/(16 T_{eff}^{2})$, and obtain the effective temperature
\bea
\frac{T_{eff}}{T} &=& \frac{1}{\sqrt{1 + 2 f_{2}(x) /(3 f_{4}(x))}}
\eea
 This expression is shown in Fig. \ref{fig:Teff}. We see that the two asymptotic values derived above are indeed visible, for $x=0$ and $x \rightarrow \infty$, and that $T_{eff}/T$ assumes a minimum in between, near $\approx 0.25$. Here, $T_{eff}/T$ is approximately $\approx 0.91$.  

\section{In-plane behavior}\label{inplanebulk}

In Eq. \label{def_ccf} we defined the in-plane current correlation function $G(\br, t)$, to quantify the behavior we have seen in Figs. \ref{fig:inplane} and \ref{fig:inplaneA0}. The full time evolution for $A_{0} = 50$ K is shown in Fig. \ref{fig:currentcorr}.
 The driving is turned on at $t = 0$ ps. After a short transient phase, the system settles into a steady state. We note that the long-range limit of $G(\br, t)$ is $V_{w}(t)$.
  We take the time average of $G(\br, t)$ in the steady state, and depict them in Fig. \ref{fig:inplanecorr}.

\end{document}